 \documentclass[12pt,preprint]{aastex}

\newcommand{\ha}{H$\alpha$}
\newcommand{\hb}{H$\beta$}
\newcommand{\har}{$\rm{H}\alpha-$$\rm{R}$}

\shorttitle{ChaMPlane CV Distribution Model}
\shortauthors{Rogel, A. B.}

\begin{document}


\title{Modeling the Galactic CV Distribution for the ChaMPlane Survey}


\author{A. B. Rogel}
\affil{Physics \& Astronomy Department, Bowling Green State University, 
    Bowling Green, OH 43402}
\author{H. N. Cohn, P. M. Lugger} 
\affil{Astronomy Department, Indiana University, 
    Bloomington, IN 47405}


\begin{abstract}
For purposes of designing targeted cataclysmic variable (CV) detection surveys 
and interpreting results of other projects with many CV detections  
such as the ChaMPlane Survey, we have created a model of the CV distribution 
in the Galaxy.  It is modeled as a warped, flared 
exponential disk with a gaussian vertical distribution.  Extinction is based on a 
detailed Galactic dust and gas model.  
A luminosity function for CVs is also incorporated, based on a smoothed version
of published data.  We calculate predicted field detection rates
as a function of the limiting magnitude 
expected for the detecting system (i.e. WIYN/Hydra or NOAO 4m/Mosaic).  
Monte-Carlo techniques are used to assess statistical fluctuations in these 
rates.  We have created maps of the expected CV 
distribution for the full non-bulge Galactic plane 
($20^\circ<l<340^\circ$, $|b|<15^\circ$) 
for use in both the ChaMPlane Survey and future CV surveys.  
Assuming a CV distribution 
with a scale height of 160 pc, the ChaMPlane observational result of
5 CVs in 13 northern fields  
is best fit by a CV local space density of $0.9^{+1.5}_{-0.5}\times10^{-5}\rm{pc}^{-3}$,
with the range representing the 95\% confidence interval.
\end{abstract}



%% \keywords{*not set yet*}


\section{Introduction}
The ChaMPlane Survey \citep{grin05} has as a primary goal the determination of the 
Galactic cataclysmic variable star (CV) distribution to check 
if CVs can provide a significant contribution to the diffuse X-ray 
background of the Galaxy.  
In the survey design, the local space density of $10^{-5}$ CVs per 
cubic parsec \citep{patt98, warn95} was used to predict 
detection rates.  Now, with
followup WIYN\footnote{The WIYN Observatory is a joint facility
of the University of Wisconsin, Indiana University, Yale University,
and the National Optical Astronomy Observatories.}/Hydra spectroscopy 
acquired for 13 fields, 11 of which are in the second and third Galactic 
quadrants \citep{roge04}, we may begin to
examine the accuracy of the published local CV space density.  
Also, we seek to make more precise predictions for expected detection 
rates along various lines of sight by using a more realistic model for
the CV distribution within the Galaxy and a detailed 
model for the gas distribution to 
compute extinction along the lines of sight.  

\section{The Model}
\subsection{Overview}
The basic Monte-Carlo process of determining CV detection rates starts 
with the creation of a simulated CV population based on adopted 
models for the Galactic CV radial distribution, vertical distribution, 
and luminosity function.
Input parameters to the model are the local CV density and the 
CV disk density vertical scale height ($h_z$), 
with assumed forms for the radial and vertical distributions 
for the CV population.

The simulated CV population consists of a Galactic position 
($l$, $b$, and distance) and absolute magnitude for each of $\sim2 \times 10^{6}$ CVs,
depending on the input parameters.  For
each field of interest, the set of CVs that lie in the field is determined.
The apparent magnitudes of these CVs are then determined based on both 
distance and the Galactic extinction derived from the adopted model.  
A limiting magnitude is applied to obtain a count of detected CVs in 
each field.  We repeat this process 1000 times to assess the 
statistical fluctuations of the results.  

\subsection{Space distribution}

The CV space distribution model is a composite model, with an exponential radial 
density profile following \citet{bm} Eq. 10.3 (disk only) giving a face-on 
surface luminosity distribution of $\Sigma\propto exp^{-R/R_{d}}$ 
with $R_{d}=2.5\rm{kpc}$.
The bulge is ignored, as we are 
primarily looking at fields in the second and third quadrants, so we will never
see a significant contribution from bulge CVs.  The two first quadrant 
fields in the \citet{roge04} sample 
are at galactic longitudes sufficiently far away from the galactic 
center to avoid the bulge as well.  This is \emph{not} the case for the
southern hemisphere ChaMPlane spectroscopy done at Magellan and CTIO, as 
those fields are near the Galactic center and thus likely include bulge CVs
\citep{zhao2005}.  
The disk model of \citet{robin2003} was followed to reproduce the 
warped and flared outer Galactic disk.

To more closely follow the work by \citet{patt84, patt98}, 
which is the source for our preliminary CV detection rates, we use a
gaussian density distribution in 
the $z$-direction \citep[][see \S V.c]{patt84},
i.e. $D\propto exp^{-[z/h_{z}]^2}$.  
CV scale heights were selected to test the full range of scale heights
suggested by \citet{patt84} of 100--250\,pc, with an additional test at 
300\,pc to cover the \citet{vanp96} suggestion that the scale height is 
a factor of 2 higher than the 150\,pc \citet{patt84} result, which was 
derived from optically detected CVs.
Note that direct comparison of scale heights between \citet{patt84} and
\citet{vanp96} is complicated by the use of a gaussian vertical
distribution by Patterson and an exponential vertical distribution
by Van Paradijs.  Our total test range for the gaussian vertical CV
distribution is thus 100-300\,pc. 

%The total number of CVs in the model is determined by the 
%local CV density and the scale height, with total CV number 
%increasing as the scale height increases.

\subsection{Luminosity distribution}

The CV luminosity distribution model is taken from \citet{patt98} 
Figure 2, with the bright supersoft binaries removed to remove 
non-accretion powered sources.  This leaves 
only the accretion-powered CV luminosity distribution.  A smoothed 
version of this distribution is used for the model CV 
population.  We adopt a color index of $V\!-\!R$ = 0.0 for all CVs.  
With this assumption, the $M_V$ given by the luminosity distribution 
is assumed equal to $M_R$ used in the apparent magnitude calculation.  
The true $V\!-\!R$ color of a CV is likely to be redder than this, 
which would result in an increase in the predicted CV detection
rate in the model. 
As the goal is to make a conservative minimum prediction of CV detections,
this choice of CV color is reasonable.  As a check of this, 
all five CV candidates and two previously-known X-ray binaries 
observed by \citet{roge04} have $V\!-\!R\!>\!0.3$, with the average $V\!-\!R\!>\!1.0$.    

The magnitudes used by \citet{patt98} to derive the luminosity function 
are time-averaged values, with the quiescent CV magnitudes being 1-2 
magnitudes fainter.  Since CVs spend most of the time in the quiescent 
state, using the time-averaged magnitudes to determine detection will result
in an overestimation of CV detections.  To eliminate this bias, the 
luminosity function was dimmed by 1.5 magnitudes to represent a 
quiescent CV luminosity function.  In tests versus a simplistic 
quiescent/outburst model (90\% quiescent, 10\% outburst of 3.8 
magnitudes brighter, equivalent to a 1.5 magnitude difference between 
quiescent and time-averaged magnitude), the use of the quiescent 
luminosity function results in about a 30\% decrease in predicted CV detections.
Versus the time-averaged luminosity function, using the quiescent 
luminosity function results in about a 70\% decrease in predicted detections.  
Clearly, the details of the CV light curve will have a significant impact 
on detection rates.  To investigate this one might draw a random sample of CV 
luminosities from a known light curve to improve the accuracy of the predicted
detections of CVs.  
However, the differing subtypes of CVs have
different outburst behavior and light curve profiles \citep{warn95}, 
necessitating an 
estimation of the Galactic distribution of each subtype of cataclysmic
varible as well
as relative numbers of each types.  This estimation would add additional
uncertainties to the model.  Thus, 
to ensure a conservative minimum prediction for CV detections, the 
quiescent luminosity function is used in this work.

\subsection{Extinction Model}

The extinction to each simulated CV is determined using the model
of \citet{drim03}, due to the full-sky coverage provided by the model.
This model uses COBE/DIRBE\footnote{DIRBE = Diffuse Infrared 
Background Experiment on NASA Cosmic Background Explorer (COBE).} 
FIR and NIR data 
to determine structure and extinction parameters of the Galactic 
dust distribution.  The model calculates $A_V$ for any location within 15 kpc of 
the Galactic core.  Regions within $20^\circ$ of the Galactic core are 
not well modeled.  However, this is not an issue in the present work, 
since all fields are at least  $26^\circ$ in longitude away from the 
Galactic center.  In all cases, the rescaling 
option included by \citet{drim03} is used to more closely match the 
fine structure of the Galactic dust and gas.  

Some authors \citep*{arce99, camb05} have suggested that the \citet{schl1998} 
data, on which the
\citet{drim03} extinction map is based, may overestimate extinction 
by a factor 
of $\sim1.3$ for regions where the extinction gradient is low and 
$A_V\gtrsim{0.8}$, while possibly underestimating extinction in regions 
where the extinction gradient is high.  We directly examine the case 
of extinction overestimation below, and see that the resultant
change in the overall predicted CV sky density is
on the order of 10-20\%, which is generally less than other 
uncertainties in the model.
The possibility of extinction 
underestimation also appears to be limited
to relatively small regions of the sky, typically around localized regions 
of significantly high extinction.  However, the localized 
underestimation seems to be of a similar order to the general overestimation,
resulting in similarly modest uncertainties in comparison to others.

\subsection{Other details}

Given the high demand for random numbers in the model (5 per CV, 
several million CVs per model, 1000 model generations), we needed 
to ensure that our numbers were properly randomized for lists 
exceeding $10^{10}$ in length.  
Therefore, we used the `Mersenne Twister' random number generator 
MT19937 \citep{mats98}\footnote{www.math.keio.ac.jp/~matumoto/emt.html}.
This generator has a total `list length' of over $10^{6400}$, so is
easily capable of generating enough random numbers for the model.

\section{Results}
\subsection{CV Sky-density Maps}
The basic model parameters are the scale height of the CV disk, the 
local CV space density, and the limiting magnitude of detection.  
For the initial CV sky-density maps, an $R$-band limiting magnitude of 22 
was selected to correspond to the expected limit for the spectroscopy
of the ChaMPlane Survey.  
A grid was laid down centered on Galactic anticenter with a grid
spacing of $0.6^\circ$ in both $l$ and $b$ to correspond to the field 
of view of the Mosaic imager 
on the NOAO 4m telescopes used in the ChaMPlane survey.  Each simulated CV was 
then placed (by sky
position) into the appropriate field, then checked against the magnitude
limit for detection.  Averages of CV detections per field were then used
to build up the maps.

\begin{figure}
\plotone{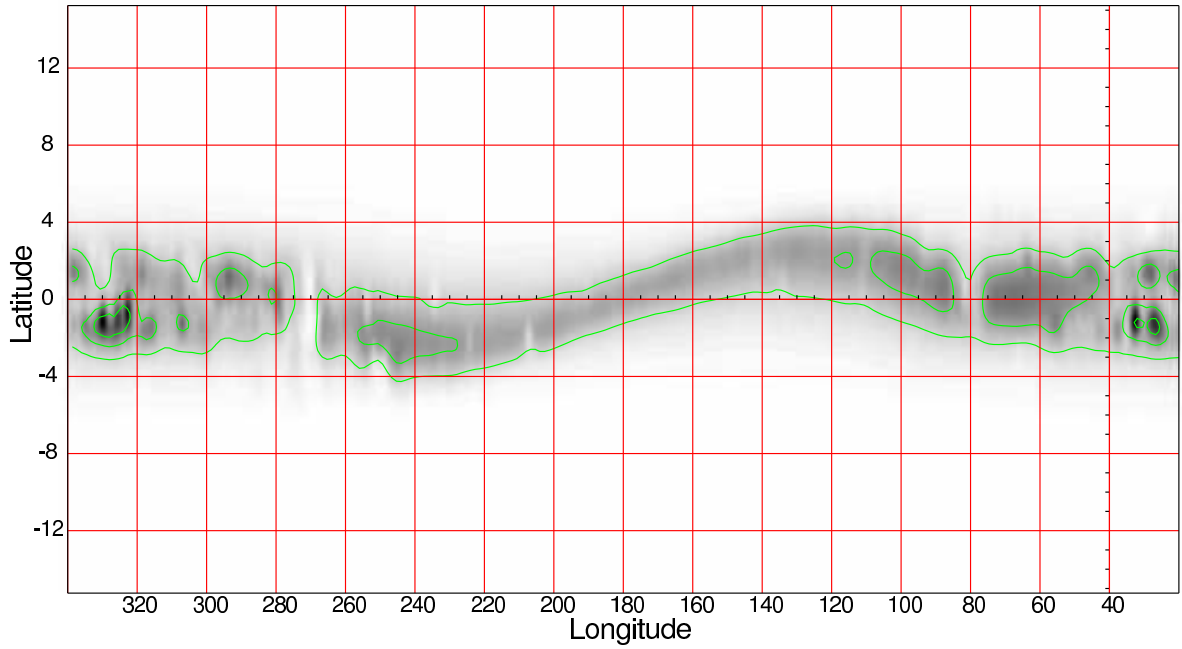}
\caption[CV detection, $R<22.0, h=100$\ pc]{Predicted CV detection rate vs.
Galactic coordinates.  $R<22.0$, scale height $h=100$\ pc.
The detection rate is in units of CV per $0.36^\circ$ square degree field.  
Contour levels start from one CV per field and are at unit intervals. 
\emph{See the electronic edition of the Journal for a color 
version of this figure.}
\label{100-220}}
\end{figure}

\begin{figure}
\plotone{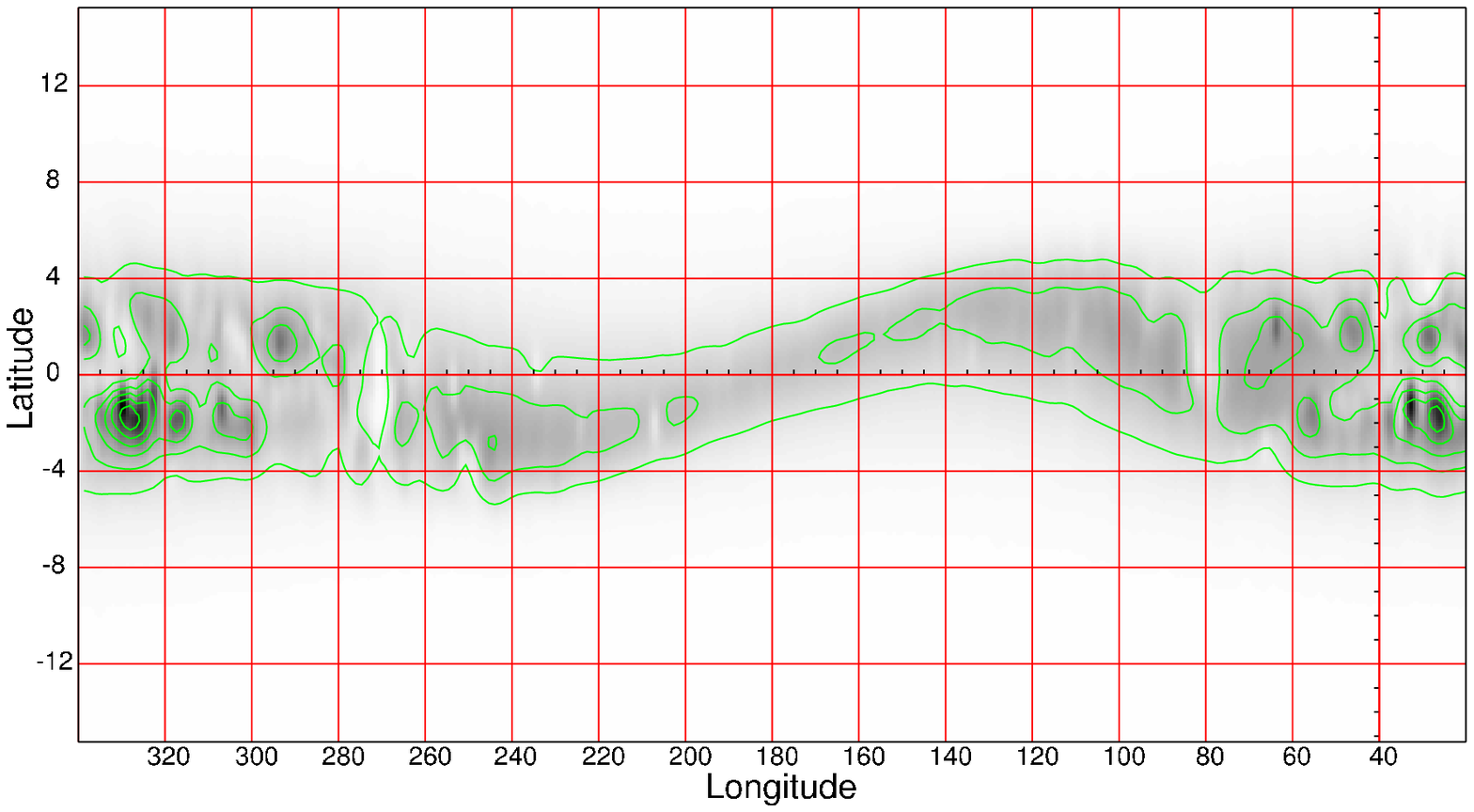}
\caption[CV detection, $R<22.0, h=160$\ pc]{Predicted CV detection rate vs.
Galactic coordinates, as for Figure \ref{100-220}.  $R<22.0$, scale 
height $h=160$\ pc.  
\emph{See the electronic edition of the Journal for a color 
version of this figure.}
\label{160-220}}
\end{figure}

\begin{figure}
\plotone{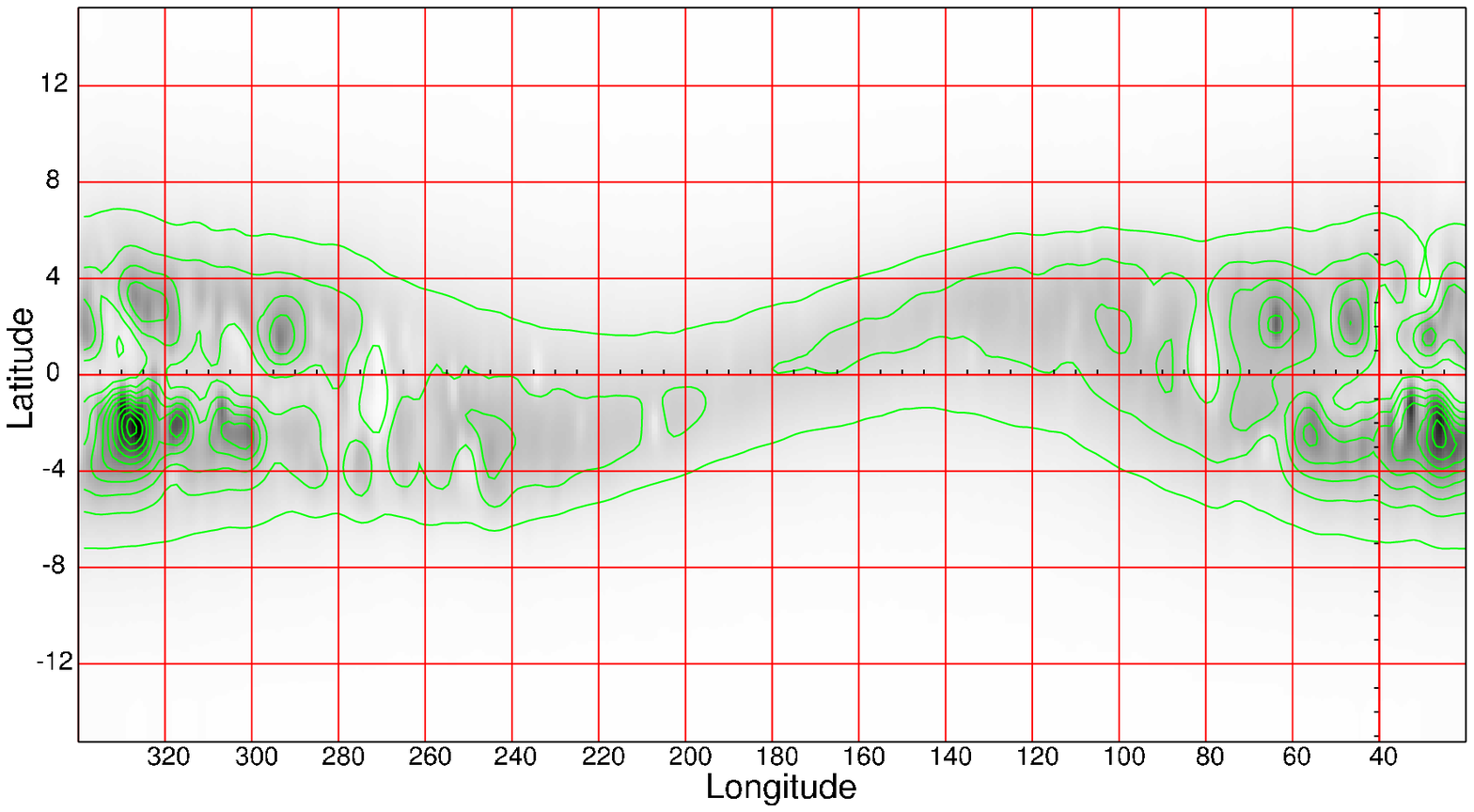}
\caption[CV detection, $R<22.0, h=230$\ pc]{Predicted CV detection rate vs.
Galactic coordinates, as for Figure \ref{100-220}.  $R<22.0$, scale 
height $h=230$\ pc.  
\emph{See the electronic edition of the Journal for a color 
version of this figure.} 
\label{230-220}}
\end{figure}

\begin{figure}
\plotone{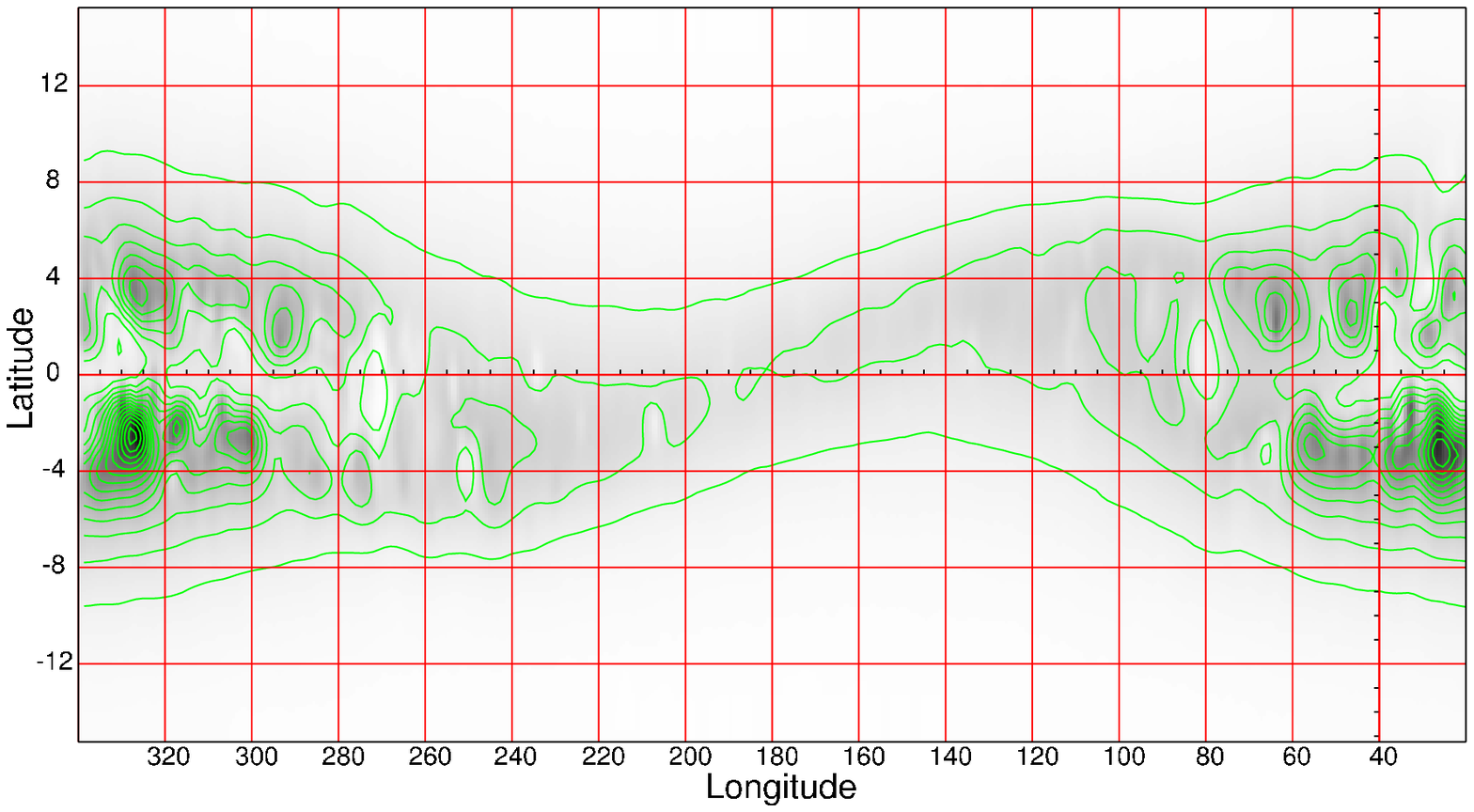}
\caption[CV detection, $R<22.0, h=300$\ pc]{Predicted CV detection rate vs.
Galactic coordinates, as for Figure \ref{100-220}. $R<22.0$, scale 
height $h=300$\ pc.  
\emph{See the electronic edition of the Journal for a color 
version of this figure.} 
\label{300-220}}
\end{figure}

Figures \ref{100-220} through \ref{300-220}
are the CV sky-density maps corresponding to four different scale heights
spanning the range discussed above.  All figures have contours set at 
unit intervals in predicted CV counts per NOAO 4m Mosaic field.  The outermost 
regions (away from the Galactic plane) all have predicted CV counts
of less that one CV per Mosaic field.  For clarity, the figures have
been stretched by a factor of five in the vertical direction.  Note also
that these figures, unlike most Galactic-coordinate sky projections, 
are centered on the Galactic \emph{anticenter}, since the region
within $20^{\circ}$\ of the Galactic center is not modeled.
All modeling runs have a local space density set at
$0.6\times10^{-5}\ \rm{pc}^{-3}$.
An important consideration in examining these figures 
is that the local space density parameter is set at the Galactic 
plane.  This means that an increased scale height (and thus thicker 
CV disk) requires \emph{more} CVs to populate, thus
resulting in generally higher CV numbers, especially away from 
the Galactic plane.  Altering the local space density results in a 
linear change in the CV sky density, so the maps may be easily scaled
to examine other local space densities.

Clearly seen in all maps is the signature of the Galactic warp and the
extinction caused by the local Orion Arm at roughly 
$l=80^{\circ}$ and $l=265^{\circ}$.  Also clear in all cases is the 
expected general increase in observed CV density as the Galactic 
longitude of the line of sight approaches the Galactic center.  The 
patchy nature of the maps is due to localized extinction features
within the Galactic plane.  In all cases, the maximum CV density is
located a few degrees (ranging from $\sim1.5^{\circ}$
for $h_z=100\ \rm{pc}$ to $\sim3.5^{\circ}$ for $h_{z}=300\ \rm{pc}$) 
below the Galactic plane,
due to the gas/dust disk of the Milky
Way being substantially thinner than 200 pc, allowing the lines of sight
a few degrees off the plane to quickly emerge from the dust disk 
while still passing through a dense portion of the CV distribution.
Only for the thickest ($h_{z}=300\ \rm{pc}$) model does the
expected observable CV density become higher than one CV per Mosaic field for
Galactic latitudes more than $\sim 9^{\circ}$\ away from the plane.  

\subsection{ChaMPlane Results}

We now apply our model to the northern ChaMPlane fields to test predicted CV detection
rates versus the 5 CV detections reported in \citet{roge04}.  For each ChaMPlane
Mosaic field, the model is run to determine a raw number of CVs predicted to
be in that field to a specified limiting magnitude.  
This number must 
then be reduced to account for survey completeness and known survey selection
effects.  

\subsubsection{ChaMPlane Survey Completeness}

Table \ref{tab-complete} lists the completion percentage for spectral 
identification of H$\alpha$-excess sources in each ChaMPlane field.  
In summary, we find that spectral identification is about 35\% complete to a 
magnitude of 20.5 for the fields.  The completion percentage (columns 5 and 8 of
Table \ref{tab-complete} is defined to be the fraction of \emph{all} (observed or not) 
H$\alpha$-excess sources with spectroscopic identifications.  
In examining Table \ref{tab-complete}, it will be noted that the fraction of
possible \ha-excess targets observed is relatively low in several cases.  This is due
to the nature of the ChaMPlane Survey, which has as its primary goal the 
characterization of the X-ray source population in the Galactic plane.  \ha-excess targets 
without known X-ray emission were given somewhat reduced priority during spectroscopy.
As the Mosaic field of view is $\sim 4.5$\ times larger than the \emph{Chandra} ACIS-I 
field of view, the X-ray emission status of most H$\alpha$-excess objects is unknown.
However, we find that observed sources have 
been identified at a success rate of 87\% to a limiting magnitude of 20.5, 
indicating both a high 
success rate at identifying sources and a significant need for further data.
For a limiting magnitude of 22.0, the total identification success rate falls to
75\%, with identification success of 55\% for objects within the magnitude
range 20.5-22.  This is attributed mainly to G-type stars which have few 
prominent spectral
features (the Na D line is heavily contaminated by skylight, and thus
is not used in our spectral identification).  We would expect to still be able 
to spectroscopically identify CVs in this dimmer magnitude range due to 
typically broad, strong \ha\ emission features.  
The five CV candidates identified by \citet{roge04} have magnitudes in the 
range 19.3 to 21.6, which supports this expectation.  The dimmest 
objects identified for each field have magnitudes ranging from 21-23. 
We therefore look at two limiting magnitude cases: one at 20.5 for 
near-certain spectral identification (which would guarantee CV identification), 
and another at 22.0 for possible CV identification.

%Over 85\% of unobserved sources lie in the 3 fields MWC297, SGR1900, 
%and M1-16.  

\subsubsection{Selection Effects}

The main selection effect in the ChaMPlane Survey to consider
is the use of a color selection
threshold of \har\ $<$\ --0.3 for potential H$\alpha$-excess emission objects, 
which corresponds to an \ha\ emission line 
equivalent width of 28\AA.  As CVs
are generally accretion-powered sources, they typically have \ha\ emission.  
To date, there is no comprehensive catalog of emission-line equivalent widths for CVs, and
even finding published values for \ha\ proved challenging (\hb\ is more commonly
cited).  However, \citet{liu1999}, \citet{will1983}, and \citet{zwit1994} each 
report \ha\ equivalent widths for a selection of CVs, obtained in studies
to spectroscopically confirm the identity of candidate CVs.  Combining
data from these three spectral surveys allows for an estimation of the fraction of
CVs with \ha\ emission line strengths below the ChaMPlane threshold value.
In total, $\sim 42\%$ of the CVs in the three surveys had equivalent widths greater 
than 28\AA, $\sim 7\%$ had multiple equivalent widths reported with values on both
sides of the cutoff, and $\sim 50\%$ had no equivalent width reported above the cutoff.
Thus, allowing for some variability as seen in multi-epoch data in the surveys, it
seems reasonable to conclude that about half of available CVs in ChaMPlane fields 
would have sufficiently strong \ha\ emission to be included as targets for 
spectroscopic identification
under ChaMPlane protocols.  An equivalent width of 28\AA\ should produce a readily-visible
feature even at faint magnitudes as discussed above.
An additional small fraction of CVs would be selected as targets based on X-ray emission,
but as the \emph{Chandra} ACIS-I field of view is $\sim 20\%$ the size of the Mosaic field, 
this will add back only a few percent of CVs lost due to low \ha\ emission, leaving
the 50\% estimate of available CVs actually being targeted by ChaMPlane spectroscopy 
essentially unchanged.

\subsubsection{Results}

\begin{figure}
\plotone{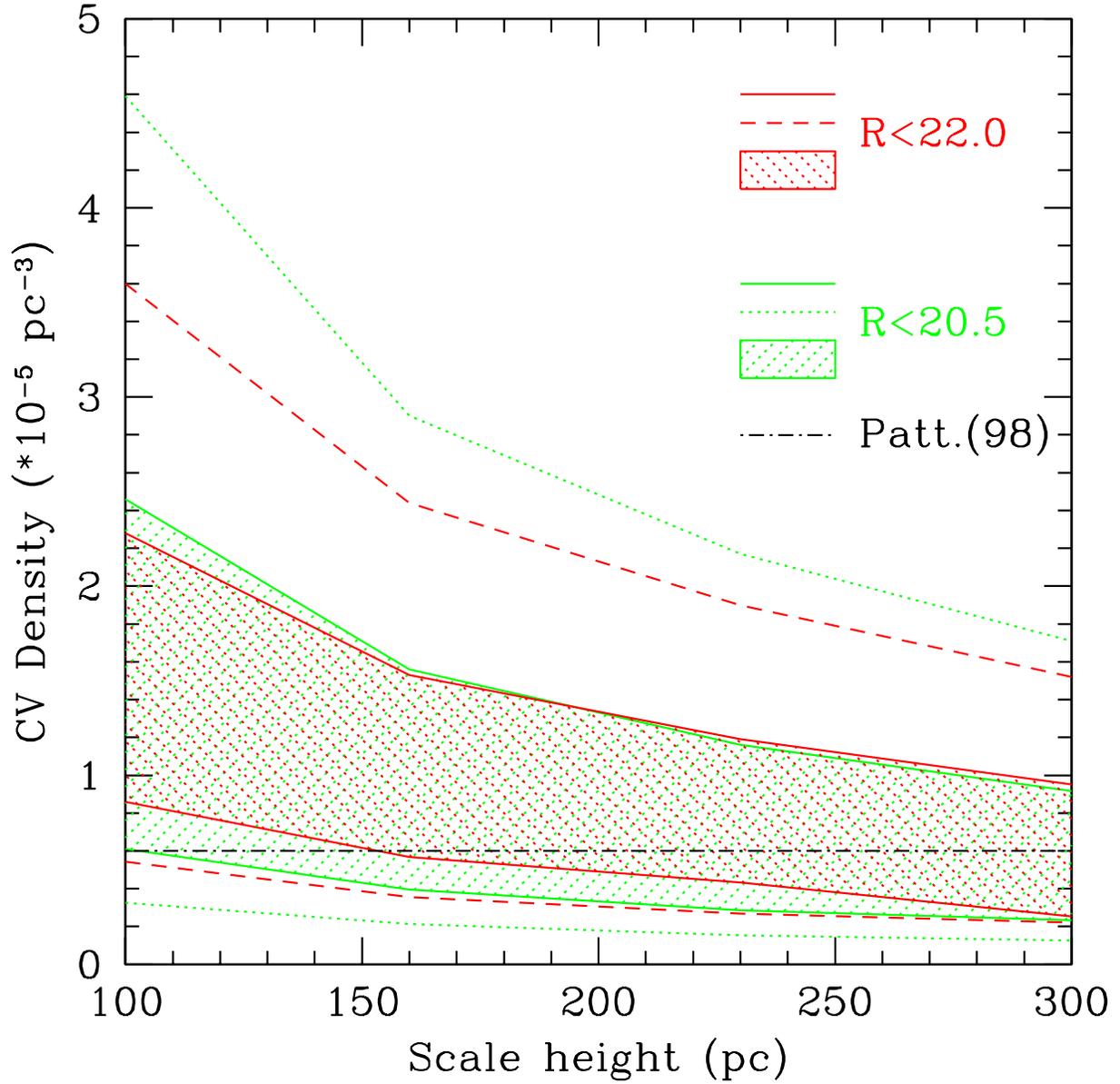}
\caption[Scale height vs. CV local space density]{Scale height vs. CV space density.
The shaded regions represent the $1\sigma$ error limit for the two different
limiting magnitudes, while the dashed/dotted lines show the $2\sigma$ error
limits.  The horizontal dot-dash line is the \citet{patt84} 
CV space density. 
\emph{See the electronic edition of the Journal for a color 
version of this figure.} 
\label{results}}
\end{figure}

Table \ref{tableone} gives the total model predictions (after taking into account
field completeness and the \ha\ line strength selection effect) 
for the 13 ChaMPlane fields with WIYN spectroscopy reported in \citet{roge04}.
Four scale heights were examined in the model, along with two limiting magnitudes.
The model was run a total of 1000 times for each field to generate the statistics.
The standard deviations quoted 
are computed directly from the Monte-Carlo results of the model for 
each field, then combined in quadrature to get a net uncertainty.
Two fields, MWC297 and SGR1900, are of concern in this work; 
SGR1900 due to short exposure times during photometry and MWC297 due 
to the presence of the Herbig-Haro Be star MWC297 and associated local 
gas and dust in the field, which is likely to produce anomalous 
extinction results as discussed by \citet{drim03}.  An examination of
the $R$-band image for MWC297 clearly shows significant ($>5$ magnitude)
extinction across more than half the image.
Included in table \ref{tableone} are results based on both all 13 ChaMPlane 
fields and for all fields but these two.  
Based on these results, and assuming a CV distribution 
with a scale height of 160 pc, the ChaMPlane observational result of
5 CVs in 13 northern fields  
is best fit by a CV local space density of $0.9^{+1.5}_{-0.5}\times10^{-5}\rm{pc}^{-3}$,
with the range representing the 95\% confidence interval.
Figure \ref{results} plots local space density versus scale height and shows
the range of possible values consistent with the ChaMPlane results to one standard deviation 
(shaded regions) and two standard deviations (lines), indicating the need for a 
higher local space density as the scale height decreases.  The \citet{patt84} 
local CV space density is consistent within two standard deviations of the ChaMPlane
results for all scale heights.  At the low end of the scale height range, the
prediction is just less than $2\sigma$ above the \citet{patt84} value.
This is also in agreement with the suggestion of \citet{grin05} that the
CV local space density derived from the \emph{ROSAT} Bright Star Survey 
\citep{schw2002} and the earlier \emph{Einstein} Galactic Plane Survey
\citep{hert1990} of $3\times10^{-5}\rm{pc}^{-3}$ is too high by a factor 
of $\sim3$.  A recent study of results from the \emph{ROSAT} North Ecliptic 
Pole survey \citep{pret07} is also in agreement with our predictions.

\subsection{Modifications to Extinction}

As noted previously, the \citet{drim03} model may overestimate extinction
in areas where $A_V$ is high.  To check the effect of this, the model
was adjusted to use the \citet{camb05} extinction modification of 
$$A_{V(\rm{Cambr\grave{e}sy})} = \frac{A_{V(\rm{Drimmel})} + 0.24} {1.31} $$
if $A_V(\rm{Drimmel})>0.774$, which is the extinction for which the 
\citet{camb05} and \citet{drim03} $A_V$ values are equal.  For comparison, the
model was run for ChaMPlane fields with scale height of 160 pc.  For the 11
true anticenter fields, the average increase in predicted CVs was 8\%.  This
is a negligible increase, considering the uncertainty in the best-fit CV 
local space density cited above.  The two quadrant 1 fields, MWC297 and 
SGR1900, had higher changes, with SGR1900 predicted CVs up by 45\% and
MWC297 predicted CVs up by 73\%.  However, the MWC297 field has a visible 
strong extinction gradient as noted above, which could lead to the 
\citet{drim03} extinction being an underestimate rather than an overestimate.
The location of the SGR1900 field at ($l=43^{\circ}, b=+0.8^{\circ}$), 
very close to the galactic plane, may 
also experience a high extinction gradient, so CV predictions 
there are possibly inflated as well.  Including these fields in the comparison
results in a 18\% increase in predicted CVs, still a small 
contribution to the total [model uncertainty]
[change when
compared to the CV local space density uncertainty.]
Other ChaMPlane fields
are not expected to be at locations with high extinction gradients, because
the regions of high extinction gradient noted in \citet{arce99}, where the \citet{drim03}
model is shown to underestimate $A_V$, are all in regions of high $A_V$.  ChaMPlane
fields were specifically selected to have low $A_V$, so problematic
locations have generally been avoided in the ChaMPlane Survey.  
A more recent extinction model, taking into account the concerns of 
\citet{camb05} and \citet{arce99}, would be beneficial for this work, in particular
for the all-sky maps.  The model of \citet{mars06} appears to avoid 
the potential shortcomings of the \citet{drim03} model, 
but as it only covers $|l|<100^{\circ}$, much of the current model's sky 
coverage is not included, including 11 of 13 ChaMPlane fields.  Thus, for 
the present application of the model to ChaMPlane results, the \citet{mars06}
model is insufficient.  However, as the CV density is clearly highest
in quadrants 1 and 4, we plan to compare 
predictions of the model
based on the \citet{drim03} extinction and a limited sky coverage version 
of the model using the \citet{mars06} extinction model  
to examine the potential effects of this extinction underestimation on 
predictions made for future targeted CV surveys.

\section{Conclusion}
Based on the model of the disk CV distribution presented here, 
it appears that the ChaMPlane Survey CV detections reported in \citet{roge04} 
for 13 northern WIYN-surveyed fields 
are consistent with a CV density of $0.6\times10^{-5}\rm{pc}^{-3}$ and the range of
scale heights proposed by \citet{patt84}.  This also supports the suggestion 
of \citet{grin05} that earlier \emph{ROSAT} and \emph{Einstein} derived 
CV space densities of 
$3\times10^{-5}\rm{pc}^{-3}$ are too high by a factor of $\sim3$.
Completion of spectroscopic identification of potential targets in the ChaMPlane
fields will result in much smaller viable ranges for the
local space density and scale heights of CVs, narrowing the allowable parameter
space.  The model will also allow future survey projects to selectively target
potentially high-output fields to significantly boost the number of new CV 
detections, thus allowing for further 
investigations of both the spatial distribution of CVs 
and also the CV luminosity function.

\acknowledgments
This research was supported in part by NSF grant AST-0098683 and
NASA/Chandra grants AR2-3002B, AR3-4002B, and AR4-5003B to Indiana University.  
Additional support was provided by NASA through the
American Astronomical Society's Small Research Grant Program.
Finally, 
ABR gratefully acknowledges a Fellowship
from the Indiana Space Grant consortium.

\clearpage

%Updated May 09 2007
\begin{deluxetable}{lcrrrrrr}
\tablecaption{ChaMPlane Field Completion Rate\label{tab-complete}}
\tablewidth{0pt}
\tabletypesize{\footnotesize}
\tablehead{
\colhead{} &
\colhead{} &
\multicolumn{3}{c}{16.0 to 20.5 mag.\tablenotemark{b}} &
\multicolumn{3}{c}{16.0 to 22.0 mag.\tablenotemark{b}} \\
\colhead{Field} &
\colhead{CV\tablenotemark{a}} &
\colhead{Targ.\tablenotemark{c}} &
\colhead{Obs.\tablenotemark{d}} &
\colhead{ID'd\tablenotemark{e}} &
\colhead{Targ.\tablenotemark{c}} &
\colhead{Obs.\tablenotemark{d}} &
\colhead{ID'd\tablenotemark{e}}
}
\startdata
3C 123 		&1     	&0 	&0 &\nodata   	& 4 	&3 &50.0\%\\
3C 129 		&0     	&3 	&1 &33.3\%    	&10 	&6 &40.0\%\\
GK Persei\tablenotemark{f} 	&0  	&6\tablenotemark{f} 	&0 &0.0\%     	&78\tablenotemark{f} 	&2 &2.6\%\\
NGC 1569 	&0   	&9 	&2 &0.0\%     	&35 	&3 &2.9\%\\
\ \ \ without core\tablenotemark{g} &0   &3 &2 &0.0\%  &4 	&2 &0.0\%\\
J2227+6122 	&0 	&11 	&10 &54.5\%   	&14 	&12 &50.0\%\\
SGR1900+14 	&0 	&109	&31 &28.4\%  	&163 	&38 &22.1\%\\
B2224+65 	&0   	&5   	&5 &100.0\%   	&11 	&7 &54.5\%\\
J0422+32 &1\tablenotemark{h} &2 &2 &50.0\% 	&8  	&8 &50.0\%\\
A0620-00 	&0   	&2     	&2 &100.0\%   	&21  	&12 &47.6\%\\
MWC297 &1\tablenotemark{h} &57 	&19 &33.3\% 	&67  	&21 &28.4\%\\
G116.9+0.2\tablenotemark{i} &1\tablenotemark{h} &11\tablenotemark{i} &4 &27.3\% &120\tablenotemark{i} &10 &4.2\%\\
PSRJ0538+2817 	&0 	&4  	&4 &50.0\%    	&19 	&9 &15.8\%\\
M1-16 		&1   	&0     	&0 &\nodata  	&83 	&4 &2.4\%\\
\enddata
\tablenotetext{a} {Number of new CV candidates in field.}
\tablenotetext{b} {$R$-band magnitude range of target list.}
\tablenotetext{c} {Total number of H$\alpha$-excess objects in field.}
\tablenotetext{d} {Number of observed H$\alpha$-excess objects in field.}
\tablenotetext{e} {Percentage of all targets (observed or not) with spectral identifications.}
\tablenotetext{f} {Many `sources' are due to the GK Persei nova shell.}
\tablenotetext{g} {Statistics without sources located in the core of NGC 1569}
\tablenotetext{h} {CV candidates with magnitudes less than 20.5.}
\tablenotetext{i} {Many `sources' are due to \ha\ filaments of a supernova remnant.}
\end{deluxetable}

%UPDATED FOR 2007!
\begin{deluxetable}{lcrrrr}
\tablecaption{ChaMPlane Field Predicted CV Detection Rates \label{tableone}}
\tablewidth{0pt}
\tabletypesize{\footnotesize}
\tablehead{
\colhead{\shortstack{Scale\\height}} &
\colhead{\shortstack{Fields\\~}} &
\colhead{\shortstack{Lim.\\Mag}} &
\colhead{\shortstack{Predicted\\Detections\tablenotemark{a}}} &
\colhead{\shortstack{Standard\\Deviation}} &
\colhead{\shortstack{Actual\\Detections}} 
}
\startdata
100 pc   &13\tablenotemark{b}       &20.5 &1.2 &1.1 &3\\
  		\nodata  &\nodata   &22.0 &2.3 &1.5 &5\\
\nodata  &11\tablenotemark{c}       &20.5 &1.0 &1.0 &2\\
		\nodata  &\nodata   &22.0 &1.7 &1.3 &4\\
160 pc   &13\tablenotemark{b}       &20.5 &1.8 &1.3 &3\\
		\nodata  &\nodata   &22.0 &3.3 &1.8 &5\\
\nodata  &11\tablenotemark{c}       &20.5 &1.5 &1.2 &2\\
		\nodata  &\nodata   &22.0 &2.6 &1.6 &4\\
230 pc   &13\tablenotemark{b}       &20.5 &2.4 &1.6 &3\\
		\nodata  &\nodata   &22.0 &4.1 &2.0 &5\\
\nodata  &11\tablenotemark{c}       &20.5 &2.1 &1.5 &2\\
		\nodata  &\nodata   &22.0 &3.4 &1.9 &4\\
300 pc   &13\tablenotemark{b}       &20.5 &2.9 &1.7 &3\\
		\nodata  &\nodata   &22.0 &5.0 &2.2 &5\\
\nodata  &11\tablenotemark{c}       &20.5 &2.6 &1.6 &2\\
		\nodata  &\nodata   &22.0 &4.1 &2.0 &4\\
%\nodata  &\nodata &22.0 &\phantom{8}6.5 &1.9 &4\\
\enddata
\tablenotetext{\ }{For a local space density of $0.6\times10^{-5}\rm{pc}^{-3}$}
\tablenotetext{a}{Factoring in completeness fraction of observations and EW(\ha) selection effect.}
\tablenotetext{b}{All 13 ChaMPlane fields.}
\tablenotetext{c}{SGR1900 field removed due to short exposures for photometry and resulting uncertainties, MWC297 field removed due to model extinction concerns due to Herbig-Haro Be star MWC297.}
\end{deluxetable}

\clearpage


\end{document}